# Using GeoGebra to discover the motion of device in a well-known physical experimental instrument

## --Looking into vibration-damping devices in the scanning tunneling microscope

Chengtian Liang[1], Enqi Xu[1]

(1. School of Physics, Hangzhou Normal University 311121, Hangzhou)

**Abstract:** In this paper, we take a vibration-damping devices in the well-known physical experimental instrument--scanning tunneling microscope as the study base, and with the help of GeoGebra software, we explain in detail the principle of damping the vibration of the damper in the magnetic field to realize the vibration-damping function of the whole device and establish a clear physical picture and a correct and comprehensive knowledge. This question also shows the process of clarifying the meaning of the problem with the help of software tools.

Currently, modern educational techniques are continuously empowering physics teaching and physics problem exploration. GeoGebra, a math software, is one of the common educational technology tools that can help the study of physics questions. It can present the process of dynamic changes in the physical model to deepen students' understanding of physical laws; it can also show the process of scientific reasoning, making the thinking method more graphic and concrete ...... All these are conducive to solving the abstract and complex problems of the physics discipline, and using intuitive visual thinking to enhance students' perceptual understanding[2], e.g.. After modeling, the question is presented first for the convenience of illustration and readers' understanding.

## 1 Presentation of the question after modeling

As shown in Fig. 1, the scanning tunneling microscope vibration-damping device consists of an insulated vibration-damping platform and a magnetic damping damper. The platform is suspended by three identical light rods symmetrically distributed about the $O'O''$ axis at the lower end $O$ of a lightweight spring, the upper end of the spring is fixedly suspended at the point $O'$, and three identical dampers symmetrically placed about *the $O'O''$* axis symmetrically placed dampers are located below the platform. As shown in Fig. 2, each damper consists of a coil fixed on the lower surface of the platform by an insulated light rod and a ferromagnet fixed on the tabletop that can generate a radial magnetic field, and the distribution of the radial magnetic field is symmetric about the center of the coil, and the magnitude of the magnetic induction at the coil is $B$. When the platform at rest is perturbed by the outside world, the coil makes a damping motion in the vertical direction in the magnetic field, and the image of its displacement change with time is shown in Fig. 3. As shown in Figure 3. It is known that the velocity at t=0 is $v_0$, the direction is downward, $t_1$, $t_2$ time amplitudes are $A_1$, $A_2$. The total mass of the platform and the three coils is $m$, the coefficient of the strength of the spring is $k$, the radius of each coil is $r$, and the resistance is $R$. When the spring deformation variable is $\Delta x$, its elastic potential energy is $\frac{1}{2}k\Delta x^2$ The elastic potential energy of the spring is $\Delta x$. Excluding air resistance, **find the law of its motion**.

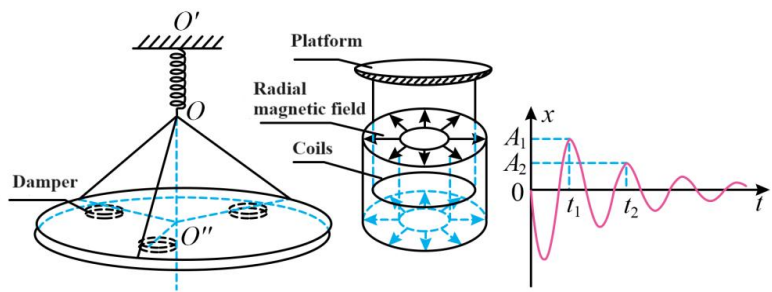

Figure 1　　　　Figure 2　　　　Figure 3

## 2 Difficulties in understanding

This question comes from modeling sets the damping system in the spring force, gravity, and amperometric resistance to do damped motion problem, the question interference and damping system of the complex form of motion - damped vibration, and directly gives the form of vibration of the approximate changes in the image as a known condition of the question. The problem-solving process can be avoided on the damped vibration displacement with time-specific functional equation, so after the problem was solved, most of the students did not have a clear understanding of the physical scenario in the problem set. In addition, all the conditions in the problem are given as physical quantities in the form of letters, which to a certain extent may cause students to have the misconception that "the specific functional equation for the change of displacement with time during damped vibration in this scenario may have nothing to do with some of these physical quantities". It is undeniable that the question "What is the specific functional relationship between displacement and time for damped motion of a vibration-damped system subjected to elasticity, gravity, and amperometric resistance" can be quite complex and difficult for teachers to answer.

Given questions are often used as the guide and the root of the problem when the exercise is prepared, the fuzzy understanding of the problem scenario will certainly affect the questions in the role of the educational function of the exercise but also is not conducive to the establishment of a clear picture of the physics of the students and a correct and comprehensive understanding of the need for this is difficult to routinely list the results of the calculation of the complex problem through effective means of interpretation[3].

## 3 GeoGebra analytical modeling
### 3.1 Theoretical Modeling Process

In this damping system, the force analysis of the damper as a whole is carried out to obtain its forces as spring's elastic force, amperometric resistance which is inverse to the velocity, and the gravity of the system. The x-axis is established with the original length of the spring at the origin of the coordinates and OO' as the positive direction, the velocity of the coil at any moment is set to be v, and the spring coefficient of strength $k = k_1$, which can be obtained:

The component force of the spring in the vertical direction $T = k_1(\Delta x_0 - x)$

Gravity of the system $G = -mg$

For the amperometric resistance to the damper as a whole, the

At any instant, the current in each coil is $I = \dfrac{B(2\pi r)v}{R} = \dfrac{2\pi rB}{R}\dfrac{dx}{dt}$

Then, at any instant, the amperometric resistance to each coil is $F = -I \times 2\pi rB = -\dfrac{B^2(2\pi r)^2 v}{R} = -\dfrac{4\pi^2 r^2 B^2}{R}\dfrac{dx}{dt}$

Therefore, the combined force in the system is $F_{tot} = G + T + 3F = -\dfrac{12\pi^2 r^2 B^2}{R}\dfrac{dx}{dt} + k_1(\Delta x_0 - x) - mg$

From Newton's second law $F_{tot} = ma = m\dfrac{d^2 x}{dt^2}$, it follows that $m\dfrac{d^2 x}{dt^2} = -\dfrac{12\pi^2 r^2 B^2}{R}\dfrac{dx}{dt} + k_1(\Delta x_0 - x) - mg$

Noting that $\Delta x_0$ is the elongation of the spring when the platform is at rest, then $k_1 \Delta x_0 = mg$, therefore:

$$m\dfrac{d^2 x}{dt^2} + \dfrac{12\pi^2 r^2 B^2}{R}\dfrac{dx}{dt} + k_1 x = 0$$

We let $\omega_0^2 = \dfrac{k_1}{m}, 2\delta = \dfrac{12\pi^2 r^2 B^2}{mR}$, then the above equation reduce to $\dfrac{d^2 x}{dt^2} + 2\delta\dfrac{dx}{dt} + \omega_0^2 x = 0$

This is a second-order linear differential equation with constant coefficients, and the form of its solution is related to the relative magnitude of the two covariates δ, ω0 in the equation. Mathematically rigorous calculations show that at this point the displacement of the whole system oscillator versus time can be expressed as:

$x = A_0 e^{-\delta t}\cos(\omega_f t + \varphi_f)$ which $\omega_f = \sqrt{\omega_0^2 - \delta^2} = \sqrt{\dfrac{k_1}{m} - \dfrac{36\pi^4 r^4 B^4}{m^2 R^2}}$

$A_0$ and φf in the above equation are determined by the initial conditions of the whole system oscillator:

$$\begin{cases} x_0 = A_0 \cos\varphi_f \\ v_0 = -\delta A_0 \cos\varphi_f - \omega_f A_0 \sin\varphi_f = -\delta x_0 - \omega_f A_0 \sin\varphi_f \end{cases}$$

So we can get that:

$$\begin{cases} A_0 = \sqrt{x_0^2 + \left(\dfrac{v_0 + \delta x_0}{\omega_f}\right)^2} \\ \varphi_f = \arctan\left(-\dfrac{v_0 + \delta x_0}{\omega_f x_0}\right) \end{cases}$$

Substituting the initial conditions for the vibrator of this damper system $x_0 = 0$, $v_0 = v_0$ the direction of the initial velocity is vertically downward, and the final is obtained:

$$\begin{cases} A_0 = \sqrt{x_0^2 + \left(\dfrac{v_0 + \delta x_0}{\omega_f}\right)^2} = \left|\dfrac{v_0}{\omega_f}\right| = -\dfrac{v_0}{\sqrt{\dfrac{k_1}{m} - \dfrac{36\pi^4 r^4 B^4}{m^2 R^2}}} \\ \varphi_f = -\dfrac{\pi}{2} \end{cases}$$

Combining the above equations, the displacement of the whole system oscillator versus time is given by:

$$x = -\dfrac{v_0}{\sqrt{\dfrac{k_1}{m} - \dfrac{36\pi^4 r^4 B^4}{m^2 R^2}}} e^{-\dfrac{6\pi^2 r^2 B^2}{mR}t} \sin\left(\sqrt{\dfrac{k_1}{m} - \dfrac{36\pi^4 r^4 B^4}{m^2 R^2}}\, t\right)$$

### 3.2 GeoGebra visualization process

(1) Create a slider $v_0$ and set the interval to (0,10), which suggests that the size of the initial velocity $v_0$ varies from 0 to 10 m/s; create a slider $k_1$ and set the interval to (0,5), which indicates that the spring strength coefficient k1 varies from 0 to 5 N/m; create a slider m and set the interval to (0,5), which suggests that the total mass of the platform and the three coils are m varies from 0 to 5 kg Create a slider r, set the interval to (0,2.5), which indicates that the radius r of each coil in the damping system varies from 0 to 2.5 m. Create a slider B, and set the interval to (0,2.5), which indicates that the magnitude of the magnetic induction B of the radial field where the coils are located in the damping system varies from 0 to 2.5 T. Create a slider R, set the interval to (0,2.5), which indicates that the magnitude of the magnetic induction B of the radiation field where the coils are located in the damping system varies from 0 to 2.5 T. R, set the interval as (0,2.5), indicating that the size of the resistance R of each coil in the damping system varies in the interval range of 0~2.5 Ω.

(2) Construct a specific function y of displacement over time for damped vibration (the pairwise relationship with the variables in the original functional equation is that t in the original equation corresponds to the present x, and x in the original equation corresponds to the present y), and enter y=if[x>=0,(-v_0)/sqrt(k_1/m-(36pi^4 r^4 B^4)/(m^2 R ^2))*e^((-(6pi^2 r^2 B^2))/(mR)*x)*sin(sqrt(k_1/m-(36pi^4 r^4 B^4)/(m^2 R^2))*x)] (NOTE: Type with "_" for subscript, "^ " for superscript, "pi" for the constant π, and "if[x>=0,]" for the function's domain of definition as x≥0)

At this point, the simulation model has been constructed. To display the variables corresponding to the horizontal and vertical coordinates, we need to create two "texts" and input "x" and "t" respectively with the checkbox "Latex formula" checked. Processed after moving to the end of the two axes at the end of the arrow, the effect is shown in Figure 4.

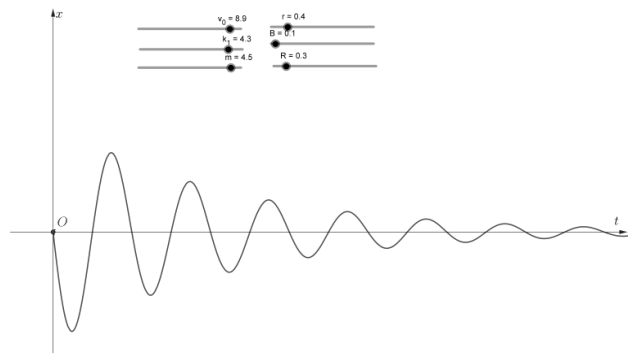

Fig. 4 GeoGebra visualization results

# 4 Analysis of GeoGebra visualization results

(1) From the visualization in Fig. 4, it can be seen that x varies approximately periodically with time t, i.e., the damper as a whole will still be vibrating near the equilibrium position, but the amplitude gradually decreases with time due to the energy reduction caused by the resistance. As the amplitude (and thus the velocity) becomes smaller, the rate of energy reduction due to the drag force becomes slower, and thus the rate of amplitude decay over time decreases[4].

(2) The damped vibration image involved in the title image is not a good fit with this model under any parameters, and can be closely and highly sensitively affected and determined by the spring stiffness coefficient $k_1$, the initial velocity $v_0$, the total mass of the platform and the three coils of m, the magnetic induction intensity B of the radial magnetic field where the coils are located in the vibration-damping system, the resistance of each coil in the vibration-damping system, R, and the radius of each coil in the vibration-damping system, r, and so on. The parameters are influenced and determined. An example is given.

Each parameter takes the following values: $v_0$ =9.1 m/s, $k_1$ =4.2 N/m, m=4.2 kg, B=0.1 T, R=0.1 Ω, the software process finds that r is only in a small range of 0~0.6 m. The graph line can be closely adapted to the situation shown in Fig. 4. When r is too large, the situation is shown in Figs. 5 and 6, and most specifically when $\dfrac{k_1}{m} \leq \dfrac{36\pi^4 r^4 B^4}{m^2 R^2}$ it is clear that the damping system fails without vibration of any kind. If the values of $v_0$, $k_1$, m, r, and R are determined, and the value of B is adjusted, the software processing finds that the value of B is also only within a small range that closely matches the damped vibration scenario involved in the title image. At the same time, the change in $v_0$ also affects the image, as shown by the overall increase in the value of x when $v_0$ is increased, i.e., the image becomes "tall and thin".

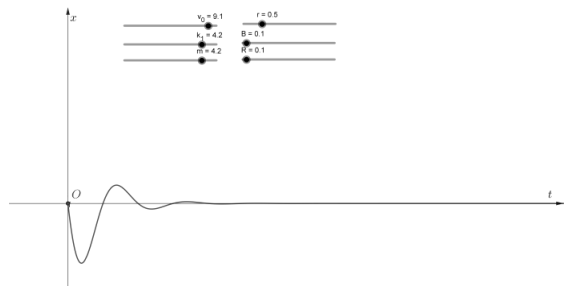

Fig. 5 Typical image when *r* is too large

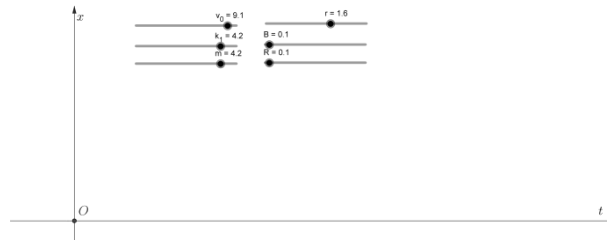

Fig. 6 Most specific image when *r* is too large ( $\dfrac{k_1}{m} \leq \dfrac{36\pi^4 r^4 B^4}{m^2 R^2}$ )

(3) When B is as small as 0, the model is simplified to a damped system only by the spring force and gravity in the case of simple harmonic motion. At this time to do simple harmonic vibration displacement x and time t into a sinusoidal function, the vibration period is controlled only by the $\dfrac{k1}{m}$ control, as shown in Figures 7 and 8.

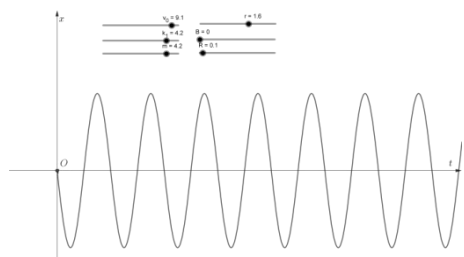

Fig. 7 Image at B=0

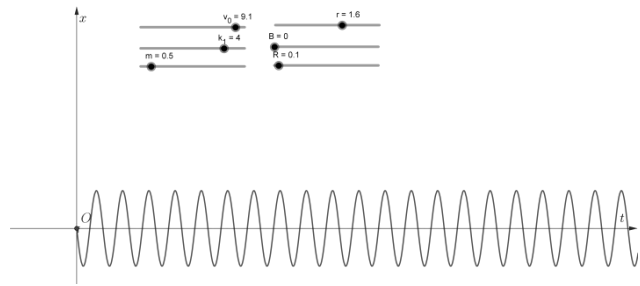

Fig. 8 The vibration period is now only controlled by the $\frac{k_1}{m}$ control

At this point, I used GeoGebra software to study the specific functional relationship between displacement and time change of the damping system in the case of damping motion by spring's elasticity, gravity, and amperometric resistance and gave some typical ranges, typical images, and the most special images of the parameters in the case of fitting with the problematic image, which can help to present the whole process of the dynamics of the physical model and deepen the students' and teachers' understanding of the physics. Laws and the connotations of puzzles.

## 5 Conclusion

In the context of the education reform, how to cope with the "new context, strong application, comprehensive and variant after the difficulty of a significant steep increase" has become secondary school teachers and students need to face a problem. In this paper, I used the mathematical software GeoGebra, the artificial difficulty to depict and visualize the function image can be displayed while achieving all the parameters can be adjusted, the effect of dynamic observation. In the face of such a variant because of the difficulty of the test questions to reach a new high, if the software tools, so that the connotation of the law of motion about vibration-damping devices can be clear.